\documentstyle[prl,aps,floats]{revtex}
\newcommand{\f}{\frac}

\begin{document}
\input epsf.sty
\twocolumn[
\hsize\textwidth\columnwidth\hsize\csname@twocolumnfalse\endcsname

\draft

\title{Correlation Lengths in Quantum Spin Ladders} 
\author{Olav F. Sylju\aa sen$^a$, Sudip Chakravarty$^a$, and Martin Greven$^b$}
\address{$^a$Department of Physics and Astronomy, University of California, 
Los Angeles, CA 90095\\
$^b$Department of Physics, Massachussetts Institute of Technology, 
Cambridge, MA 02139}
\date{\today} 
\maketitle

\begin{abstract}
Analytic expressions for the correlation length temperature dependences are given
for antiferromagnetic
spin--1/2 Heisenberg ladders using a finite--size non--linear $\sigma$--model
approach. These calculations rely on identifying three
successive crossover regimes as a function of temperature. In each of these
regimes, precise and controlled approximations are formulated. The
analytical results are
found to be in excellent agreement with Monte Carlo simulations for
the Heisenberg Hamiltonian.
\end{abstract}
]

Low--dimensional quantum magnets exhibit many novel collective 
low--temperature properties. 
An interesting recent development is the system
of spin ladders \cite{experiments} that are 
arrays of $n$ coupled antiferromagnetic Heisenberg chains. 
Experiments and numerical studies have revealed many fascinating aspects of these systems, such as the surprise 
that spin--1/2 ladders composed of an even number of chains have a 
gap in the excitation spectrum, 
while those with an odd number of chains are gapless \cite{Science}. 

For the two--dimensional square--lattice  spin--1/2 Heisenberg antiferromagnet with nearest--neighbor interactions, theoretical predictions for the spin--spin correlation length based on the (2+1)--dimensional non--linear $\sigma$--model \cite{CHN} are in excellent quantitative agreement with neutron scattering results \cite{NEUTRONS}.
In Ref. \cite{CHA}, one of us pointed out that it is useful to view 
a spin ladder as a finite--sized two--dimensional antiferromagnet,
instead of a system of coupled spin chains. 
The zero--temperature properties could then be readily
obtained from the finite--size scaling of the (2+1)--dimensional non--linear
$\sigma$--model, where one of the spatial directions is finite and the remaining
directions infinite, one of them being the Euclidean time direction. In this
manner one could obtain both a clear picture of the dimensional crossover to
the two--dimensional square--lattice antiferromagnet, as well as precise analytical
estimates of the zero--temperature correlation lengths and gaps in spin ladders. 
Moreover, it was predicted
that the spin gap,
$\Delta$, is simply related to the spin--spin correlation length, $\xi$, such that
\begin{equation}   
    \left( \f{\Delta}{J} \right)
    \left( \f{\xi}{a} \right) = 
    \left( \f{\hbar c}{Ja} \right),
\label{zero-Tgap}
\end{equation}
where $J$ is the antiferromagnetic coupling, $a$ is the lattice spacing, and
$c$ is the physical spin wave velocity of the 
{\em two--dimensional} square lattice antiferromagnet at zero temperature.
Applied to $S=1/2$, the right hand side of Eq. (\ref{zero-Tgap})
should be 1.68. 
 
At first sight, an analysis at non--zero temperatures similar to that at zero temperature may appear formidable. One of the purposes of this paper is to show that this is not so. Once one recognizes three distinct crossover scales as a
function of temperature, the temperature dependence of the correlation length can be computed with equal ease.  
In the language of the $\sigma$--model, the system is a box whose two sides $L_y=na$ (the spatial direction along the width of the ladders) and $\beta \hbar c$ (the Euclidean time direction) are finite in extent, where $\beta$ is the inverse temperature. 
Successive crossovers take place as the system goes through regimes where $\beta\hbar c \gg L_y$, $\beta\hbar c \sim L_y$, and $\beta \hbar c \ll L_y$. (There is an additional crossover at even higher temperatures, where the system becomes entirely classical, which is not discussed here.) 
In each of these three regimes, it is possible to formulate precise and controlled analytical methods to compute correlation lengths of spin ladders. 
These estimates are then tested
against new numerical simulations, and we find excellent agreement 
between the numerical and the analytical results. 
Because we use periodic
boundary conditions across the width of the ladders in the
$\sigma$--model approach \cite{Martin-Delgado}, 
it is necessary for us to carry out simulations
with periodic boundary conditions as well. 
However, we show that the choice
of boundary conditions hardly matters for ladders of width larger than $n=4$. 
Finally, we show that the zero--temperature
predictions of the correlation lengths \cite{CHA,Sierra2}, that contain no adjustable
parameters, are in good agreement with the simulation results,
and that the relation shown in Eq.~(\ref{zero-Tgap}) holds with remarkable accuracy.
Thus, we believe that we have a reasonably complete picture of spin fluctuations
in spin ladders.

\begin{figure}[htb]
\epsfxsize=3.4in 
\epsfbox{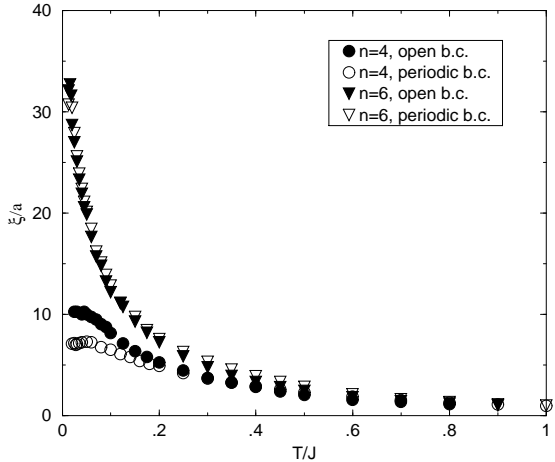}
\caption{Correlation lengths obtained from Monte Carlo simulations as 
functions of temperature for spin--1/2 ladders composed of four and six chains.}
\label{Monte}
\end{figure}

The Hamiltonian for a Heisenberg ladder is
\begin{equation}
H = J_{\|} \sum_{\langle ij\rangle_{\|}} {\bf S}_i \cdot {\bf S}_j
  + J_{\bot} \sum_{\langle ij\rangle_{\bot}} {\bf S}_i \cdot {\bf S}_j,
\label{H}
\end{equation}
where ${\bf S}_i = \frac{1}{2}{\bf\sigma}_i$ is the quantum spin operator
at site $i$, while $\langle ij \rangle_{\|}$ and
$\langle ij \rangle_{\bot}$ denote nearest neighbors along and between chains,
respectively.
The couplings considered are isotropic and antiferromagnetic, that is,
$J_{\|} = J_{\bot} = J$ and $J > 0$.
As in the previous study \cite{Greven}, the ladders are investigated 
with a very efficient loop cluster algorithm \cite{ELM-WY}.
The correlation length is obtained from the exponential large--distance 
decay of the measured instantaneous spin--spin correlation function.
We refine the original simulations with open boundary
conditions \cite{Greven}, and repeat them with periodic boundary conditions.
These calculations yield low--T correlation
lengths, $\xi/a$, of 7.1(1) and 10.3(1) for ladders of width $n=4$ for periodic and
open boundary conditions, respectively. The results for 
$n=6$ are 30.5(10) and
32.0(10) for periodic and open boundary conditions, respectively. Our corresponding
analytical results for the periodic boundary condition are 6.2 and 26.2 for 
$n=4$ and $n=6$, respectively\cite{CHA,Sierra2}. 
The spin gaps are obtained 
from independent numerical measurements of the uniform susceptibility.
For $n=4$ they are 0.234(4) and 0.160(4) for periodic and open
boundary conditions, respectively. Similarly, for $n=6$, we obtain 
0.055(4) and 0.053(4), respectively, for periodic and open boundary conditions. It is
seen that Eq.~(\ref{zero-Tgap}) holds remarkably well. The temperature dependence
of the correlation lengths obtained from Monte Carlo simulations are shown in Fig.
(\ref{Monte}). 
For $n=6$, the choice of boundary conditions indeed hardly matters.

By studying the effective Euclidean action of the (2+1)--dimensional non--linear $\sigma$--model,
\begin{equation} \label{2+1}
 \f{S}{\hbar} =  \f{\rho_s^0}{2\hbar} \int_0^{\beta \hbar} d\tau
 \int dx \int_0^{L_y} dy \left[ \left( c^{-1} \partial_\tau \hat{\Omega}
                            \right)^2 + 
                     \left( \partial_\mu \hat{\Omega} \right)^2
                     \right],
\end{equation}
where $\mu$ is to be summed over the spatial directions, $x$ and $y$,
we calculate the temperature dependence of the correlation length
in three distinct regimes characterized by the ratio $\beta \hbar c / L_y$.
The parameter $\rho_s^0$ is the bare spin stiffness constant at the spatial cutoff $\Lambda^{-1}$ of the model, $c$ is the spin--wave velocity 
defined on the same scale and $\hat{\Omega}$, the staggered
order parameter field, is a three-component unit vector.
In  the low--T ($\beta \hbar c \gg L_y$) and high--T ($\beta \hbar c \ll L_y$)
regimes we map 
(\ref{2+1}) to a (1+1)--dimensional quantum nonlinear $\sigma$--model where 
the spatial dimension refers to the $x$--direction, and the extent of the
``time--direction" is $\beta \hbar c$ and $L_y$, respectively. For the intermediate--T ($\beta \hbar c \sim L_y$)
regime the effective model is the one--dimensional classical
non--linear $\sigma$--model. The fluctuations that are integrated out
are approximated by a one--loop modification of the 
coupling constant for the effective theory.
\begin{figure}
\epsfxsize=3.4in 
\epsfbox{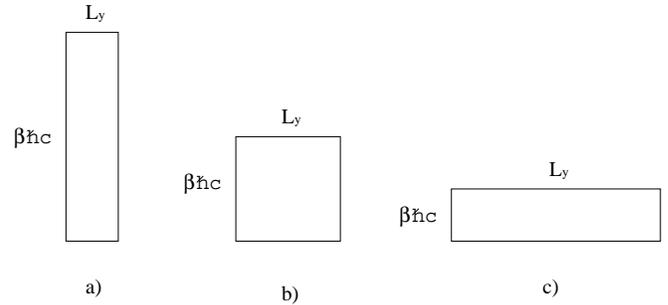}
\caption{The three different temperature regimes:
 a) low--T, b) intermediate--T, and c) high--T. The figures show the extensions 
of the two finite directions. For a) and c),  the effective model is 
a (1+1)--dimensional quantum non--linear $\sigma$--model with fields constant in the shortest directions,
and a coupling modified by the fluctuations in this direction. For b), the
fields are constant along both the $y$-- and $\tau$--directions, 
and the effective model is a one--dimensional classical $\sigma$-model.}
\label{box}
\end{figure} 
Figure (\ref{box}) illustrates the situations in the three different regimes.

At $T=0$ the ``Lorentz"--invariance of the non--linear $\sigma$--model gives
the simple relation Eq. (\ref{zero-Tgap}) between the correlation length and
the gap. For $T \neq 0$ it is not so clear that the spin--wave velocity
will survive quantum fluctuations unrenormalized. However, it is certainly
the case that the quantum fluctuations at the {\em one--loop} level cannot
renormalize the spin--wave velocity at any temperature. Therefore,
within our one--loop approximation, it is consistent to promote
Eq. (\ref{zero-Tgap}) to be valid at all temperatures, which we will do.

For low temperatures, the same procedure as that used
for the $T=0$ calculation\cite{CHA}, of integrating
out the fluctuations along the finite--width ($y$--direction) to 
one--loop order, should work.
However, at non--zero temperatures the resulting effective model is not 
the two--dimensional classical non--linear $\sigma$--model, 
but a (1+1)--dimensional quantum non--linear $\sigma$--model, 
\begin{equation}  \label{lowgap}
   \f{S}{\hbar} = \f{1}{2 \epsilon(L_y,T)} \int dx \int_0^{\beta \hbar c} du
             \left[ \left( \partial_u \hat{\Omega} \right)^2
                   + \left(\partial_x \hat{\Omega} \right)^2 \right],
\end{equation}
where
\begin{eqnarray}
   \f{1}{\epsilon(L_y,T)} &=& \f{1}{\epsilon (L_y,0)}\nonumber \\
                        &-&\int \f{dk}{2\pi}
                        \sum_{m=-\infty}^{\infty \; \; \prime}
                        \f{\left[ k^2+k_m^2 \right]^{-1/2}}
                      {\exp{\left[ \beta \hbar c \Lambda 
                            \sqrt{k^2+k_m^2} \right]}-1}. 
\end{eqnarray}
The effects of the fluctuations in the
$y$--direction are taken into account by modifying the coupling constant to
one--loop order; for convenience, we separate out the 
temperature--independent part, $\epsilon(L,0)$. In the
above expression 
$k_m=2\pi m/L\Lambda$, and the prime on the summation sign means that the
$m=0$ term is omitted from the sum. This expression is valid for ladders
with periodic boundary conditions in all directions. In
this Letter we only consider spin ladders with these boundary
conditions. 
A short wavelength cutoff is imposed on the integration and on the summation
such that $\sqrt{k^2+k_m^2} < 1$, where $\Lambda$ is the cutoff scale chosen 
to conserve the area of the first Brillouin zone: 
$\Lambda a = 2 \sqrt{\pi}$.
The results are only weakly dependent on this choice.
Because of the absence of the topological term, the above action
only describes spin--1/2 ladders of even width.

The correlation length for the above model (\ref{lowgap}) 
can be obtained from a simple self--consistent 
approach. By relaxing the unitarity condition on $\hat\Omega$ and adding 
a mass term, the condition $<\Omega^2 >=1$ yields  
\begin{equation}
   N \epsilon(L_y,T) \int_{0}^{1} \f{dk}{2 \pi} 
   \f{\coth{\left( \sqrt{k^2+ \tilde{\Delta}^2(L_y,T)} \beta \hbar c \Lambda/2 \right)}}
     {\sqrt{k^2+\tilde{\Delta}^2(L_y,T)}} =1,
\end{equation}
where $N=3$ and $\tilde{\Delta}=(\Lambda \xi)^{-1}$. 
Solving this equation at $T=0$ gives
\begin{equation}
    \tilde{\Delta}(L,0) = \f{1}{\sinh{\left(2\pi /N \epsilon(L_y,0)\right)}}.
\end{equation}

\begin{figure}
\epsfxsize=3.7in
\epsfbox{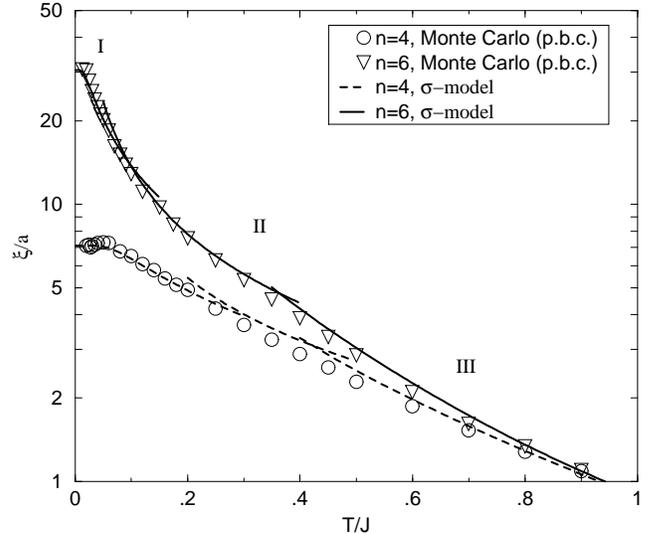}
\caption{Correlation lengths as functions of temperature in the three
distinct regimes for spin ladders of width $n=4$ and $n=6$, compared with the
data from numerical simulations. The three
regimes are labeled: I) low--T, II) intermediate--T, 
and III) high--T.}
\label{anal}
\end{figure}

Unfortunately this is not the correct zero--temperature result\cite{CHA,Sierra2}. To
remedy this we can let $N \epsilon(L_y,0)$ be a free parameter and 
adjust it such that $\Delta(L_y,0)$ agrees with the correct zero--T
result\cite{CHA}. 
Fixing $N\epsilon(L_y,0)$ in this way
we can solve Eq. (6) numerically for the correlation length at non--zero
temperatures.  
Due to the slight discrepancy between the $T=0$ correlation
lengths given in \cite{CHA} and the correlation lengths obtained in the Monte
Carlo simulations we adjust $N \epsilon(L_y,0)$ such that the
$T=0$ correlation length agrees with the Monte Carlo
data, that is $\xi/a$=7.1 and 30.5 for $n=4$ and $n=6$, 
respectively. The finite--T correlation lengths for these
ladders are shown in region I in Fig. (\ref{anal}).

For intermediate temperatures, $\beta \hbar c \sim L_y$, 
the fluctuations in both the $y$-- and the $\tau$--directions are integrated
out. The low--energy effective action is then a one--dimensional 
classical non--linear $\sigma$--model,
\begin{equation} \label{momentum}
     \f{S}{\hbar} = \f{1}{2 t(L_y,T)} \int dx 
                    \left(\partial_x \hat{\Omega} \right)^2,
\end{equation}
where the coupling constant to one--loop order is
\begin{equation}  \label{dobbelsum}
    t^{-1}(L_y,T) = \beta L \rho_s^0 -\int \f{dq}{2\pi}
                  \sum_{m,n}^{\; \; \prime}
                  \f{1}{q^2+q_m^2+\omega_n^2}.
\end{equation}
Here, $q_m=2\pi m/L_y$, $\omega_n=2\pi n/ \beta \hbar c$, and
the prime on the sum indicates that the $m=n=0$ term is omitted.
Using a combination of Ewald and Poisson 
summation techniques, as in Ref. \cite{RGJ}, we get
\begin{equation}
   t^{-1} (L_y,T)  = \f{L_y \beta \hbar c}{\xi_J} + L_y A(\f{\beta \hbar c}{L_y}),
\end{equation}
where $\xi_J$ 
is the Josephson length in the N\'eel phase\cite{CHA}, 
and the scaling function $A$ is
\begin{equation}
   A(x)=\f{\sqrt{x}}{2 \pi} \left[ \int_1^{\infty} dy y^{-1/2} \left(
                   X(\pi x y) X(\pi y /x) -1 \right)-2 \right],
\end{equation}
where 
\begin{equation}
   X(y) = \sum_{n=-\infty}^{\infty} e^{-y n^2}.
\end{equation}
 
The one--dimensional classical $\sigma$--model is nothing but the
continuum limit of the classical Heisenberg spin chain on a lattice,
which has the action
\begin{equation}    \label{lattice}
  S = -K  \sum_{i} \hat{\Omega}_i \cdot \hat{\Omega}_{i+1}.
\end{equation}
The correlation length for this lattice model has been
calculated exactly in \cite{Fisher}, and is
\begin{equation}   \label{Fisherexact}
  \f{\xi}{a} = - \f{1}{\ln \left[ \coth K - 1/K \right]}.
\end{equation}
Taking the continuum limit $a \to 0$,
with $Ka$ kept fixed, we find
\begin{equation}    \label{latticecont}
    S = \f{Ka}{2} \int dx \left( \partial_x \hat{\Omega} \right)^2 + const.
\end{equation}
The one--dimensional non--linear $\sigma$--model is a finite theory
for which no ultraviolet regularization is necessary, and there are no
ambiguities associated with taking the continuum limit.
Thus from Eqs. (\ref{momentum})
and (\ref{latticecont}) we can identify $Ka=1/t(L_y,T)$. 
Taking the continuum limit of Eq. (\ref{Fisherexact}), with $Ka$ held fixed,
we obtain
\begin{equation}
    \f{\xi}{a} = \f{t^{-1}(L_y,T)}{a}.
\end{equation}  
We note that this expression has {\em no adjustable parameters}.
The results for ladders composed of four and six chains are shown in region II
of Fig.~(\ref{anal}).

For temperatures such that $L_y \gg \beta \hbar c$, it is reasonable to 
map the action Eq. (\ref{2+1}) to a (1+1)--dimensional quantum non--linear
$\sigma$--model in which the ``time direction" is really the width $L_y=na$
of the ladders,
\begin{equation}
 \f{S}{\hbar} = \f{1}{2 f(L_y,T)} \int dx \int_0^{L_y} d y
             \left[ \left(\partial_x \hat{\Omega} \right)^2
                   + \left(\partial_y \hat{\Omega} \right)^2 \right].
\end{equation}
Here, the effective coupling constant is
\begin{eqnarray}
   \f{1}{f(L_y,T)} &=& \f{1}{f(\infty,T)}\nonumber \\
 &-&\int \f{dk}{2\pi} 
   \sum_{n=-\infty}^{\infty \; \; \prime} 
   \f{\left[ k^2+\omega_n^2 \right]^{-1/2}}
     {\exp{\left[ L_y \Lambda \sqrt{k^2+\omega_n^2} \right]} -1}, 
\end{eqnarray}
and $\omega_n=2\pi n/\beta \hbar c \Lambda$.
This is very similar to the low--T case; one just
has to interchange $\beta \hbar c$ with $L_y$. In the low--temperature 
regime, we imposed a cutoff in the two spatial directions. Here, it is
convenient to impose the cutoff in the $x-\tau$--plane. 
We have checked that this
change in the cutoff procedure does not alter the results. 
In the same way as for the low--temperature 
regime we find the correlation length by
a self--consistent equation,
\begin{equation}   \label{gap-eqL}
   N f(L_y,T) \int_0^1 \f{dk}{2\pi} 
   \f{\coth{\left( \sqrt{k^2+\tilde{\Delta}^2(L_y,T)} L_y \Lambda/2 \right)}}{\sqrt{k^2+\tilde{\Delta}^2(L_y,T)}}
   = 1.
\end{equation}
The quantity $N f(\infty,T)$ is adjusted such that the solution of the  
equation at $L_y=\infty$, $\xi (\infty,T)$, corresponds to the 
correlation length of the (2+1)--dimensional quantum
 antiferromagnet which was calculated in \cite{CHN,HAS},
\begin{equation}
  \f{\xi( \infty,T)}{a} = \f{e \sqrt{2} Z_c}{8 \pi S Z_{\rho_s}}
                       e^{2\pi S^2 Z_{\rho_s} J/T}
                       \left( 1 - \f{1}{4 \pi S^2 Z_{\rho_s}} \f{T}{J} \right),
\end{equation}
where $Z_c=1.19$ and $Z_{\rho_s}=0.740$\cite{Beard}.
Having fixed $Nf(\infty,T)$, we calculate $\xi(L_y,T)$ by solving
(\ref{gap-eqL}) numerically.
The results for $n=4$ and $n=6$ correspond to region III
of Fig.~(\ref{anal}).

It is evident from Fig. (\ref{anal}) that the analytic results for the
correlation lengths agree remarkably well with the numerical simulations
on even--legged spin--1/2 ladders with periodic boundary conditions. 
Since our effective non--linear $\sigma$--model does not contain the topological
term, our results are strictly valid for even--leg ladders.
However, the topological term is important only for the
physics\cite{CHA} at very low energies. Thus, we expect that our expressions for
the intermediate-- and high--temperature regimes are valid also for 
odd--leg ladders. It would, however, be interesting to have a more quantitative
understanding of the crossover to the WZW--model for odd--legged spin ladders.

O.F.S. and S.C. were supported by a grant from the National Science Foundation,
Grant No. DMR-9531575.
M.G. would like to thank R. J. Birgeneau and U.--J. Wiese for many 
stimulating discussions, and was supported by the MRSEC Program of 
the NSF under Award No. DMR 94-00334 and by the International Joint
Research Program of NEDO (New Energy Industrial Technology Development 
Organization, Japan).


\begin{references}
\bibitem{experiments} 
            D. C. Johnston {\em et al.}, Phys. Rev. B {\bf 35}, 219 (1987);
            Z. Hiroi {\em et al.}, J. Solid State Chen. {\bf 95}, 230 (1991);
            M. Azuma {\em et al.}, Phys. Rev. Lett. {\bf 73}, 3463 (1994);
            K. Kojima {\em et al.}, Phys. Rev. Lett. {\bf 74}, 2812 (1995).
\bibitem{Science} For a review, see 
                  E. Dagotto and T. M. Rice, Science {\bf 271}, 618 (1996)
                  and references therein.
\bibitem{CHN} S. Chakaravarty, B. I. Halperin, and D. R. Nelson, 
               Phys. Rev. B {\bf 39}, 2344 (1989).
\bibitem{NEUTRONS} M. Greven et al., Z. Phys. B {\bf 96}, 465 (1995).
\bibitem{CHA} S. Chakravarty, Phys. Rev. Lett. {\bf 77}, 4450 (1996).
\bibitem{Martin-Delgado} M. A. Martin-Delgado and G. Sierra  (private
        communications) have investigated the case of Dirichlet boundary
        condition.
\bibitem{Sierra2}See also, G. Sierra, preprint, cond-mat/9610057.
\bibitem{Greven}M. Greven, R. J. Birgeneau, and U. -J.  Wiese,
                Phys. Rev. Lett. {\bf 77}, 1865 (1996).
\bibitem{ELM-WY}H. G. Evertz, G. Lana, and M. Marcu,
                Phys. Rev. Lett. {\bf 70}, 875 (1993);
                U.--J. Wiese and H.--P. Ying,
                Z. Phys. B {\bf 93}, 147 (1994).
\bibitem{RGJ} J. Rudnick, H. Guo, and D. Jasnow, J. Stat. Phys. 
              {\bf 41}, 353 (1985).
\bibitem{Fisher} M. E. Fisher, Amer. Jour. Phys. {\bf 32}, 343 (1964).
\bibitem{KVE} D. Khveshchenko, Phys. Rev. B {\bf 50}, 380 (1994); 
              G. Sierra, J. Phys. A {\bf 29}, 3299 (1996).
\bibitem{HAS} We use the prefactor of P. Hasenfratz and F. Niedermayer, Phys.
              Lett. B {\bf 268}, 231 (1991).
\bibitem{Beard} R. R. P. Singh, Phys. Rev. B {\bf 39}, 9760 (1989);
                R. R. P. Singh and D. Huse, Phys. Rev. B {\bf 40}, 7247 (1989)
                have given the values for arbitrary $S$. Applied to
                $S=1/2$ their values are $Z_c=1.18$ and $Z_{\rho_s}=0.724$. 
                For $S=1/2$ we use the values given in 
                B. B. Beard and U.--J. Wiese, Phys. Rev. Lett. {\bf 77}, 5130
                (1996), which is $Z_c=1.19$ and $Z_{\rho_s}=0.740$.
\end{references}
\end{document}